\documentclass{webofc}
\usepackage[varg]{txfonts}   

\newcommand{\beq}{\begin{equation}}
\newcommand{\eeq}{\end{equation}}
\newcommand{\bea}{\begin{eqnarray}}
\newcommand{\eea}{\end{eqnarray}}

\begin{document}
\title{Analysis of Static Wilson Line Correlators from Lattice QCD at Finite Temperature with $T$-matrix Approach}
%
%

\author{\firstname{Zhanduo} \lastname{Tang}\inst{1}\fnsep\thanks{\email{zhanduotang@tamu.edu}} \and
        \firstname{Swagato} \lastname{Mukherjee}\inst{2}\fnsep\thanks{\email{swagato@bnl.gov}} \and
        \firstname{Peter} \lastname{Petreczky}\inst{2}\fnsep\thanks{\email{petreczk@bnl.gov}} \and
        \firstname{Ralf} \lastname{Rapp}\inst{1}\fnsep\thanks{\email{rapp@comp.tamu.edu}}
}

\institute{Cyclotron Institute and Department of Physics and Astronomy,
Texas A\&M University, College Station, TX
  77843-3366, USA
\and
           Physics Department, Brookhaven National Laboratory, Upton, New York 11973, USA }

\abstract{We employ a thermodynamic $T$-matrix approach to study Wilson line correlators (WLCs) for a static quark-antiquark pair within the quark-gluon plasma (QGP). By refining earlier determined input parameters, self-consistent $T$-matrix calculations
can approximately reproduce the WLCs from 2+1-flavor lattice-QCD (lQCD) computations, while also satisfying constraints on the bulk medium from the QGP equation of state.
The main difference from earlier results is a reduced screening in the input potential toward higher temperatures. When elaborating the consequences of this scenario on transport properties we find a more pronounced temperature dependence of the low-momentum relaxation rate of heavy quarks in the QGP. This, in particular, leads to a rather weak increase of the (temperature-scaled) spatial diffusion coefficient, which turns out to be in fair agreement with the most recent pertinent lQCD data.
}
\maketitle
\section{Introduction}
\label{sec_intro} 
The study of quark-gluon plasma (QGP) involves complex phenomena that arise from the fundamental interactions in Quantum Chromodynamics (QCD). Heavy-flavor (HF) particles can serve as versatile probes for understanding QGP properties in ultrarelativistic heavy-ion collisions (URHICs): their large masses enable potential approximations and a Brownian motion description in the QGP, and lead to a prolonged thermalization time that creates a sensitivity to their interaction history~\cite{Prino:2016cni,Rapp:2018qla,Dong:2019unq}. Quarkonia, the bound states of a heavy quark and antiquark, offer further insights into QGP properties. Lattice-QCD (lQCD) computations provide first-principles information about in-medium properties of quarkonia, such as heavy-quark (HQ) free energies and Euclidean correlators~\cite{Petreczky:2008px,Ding:2012sp}, although there interpretation is usually not straightforward. In the present paper we focus on Wilson line correlators (WLCs) for a static quark-antiquark pair at finite temperature. Recent 2+1-flavor lQCD computations (with realistic pion masses) have found that the computed WLCs cannot be described by hard-thermal loop perturbation theory predictions~\cite{Bala:2021fkm}, while the use of various fit functions suggested a rather weak screening of the underlying potential along with large spectral widths. Here, we deploy an in-medium $T$-matrix approach that is nonperturbative in nature and allows to perform microscopic calculations of HQ properties in a strongly coupled QGP~\cite{Mannarelli:2005pz,Cabrera:2006wh,Riek:2010fk,Riek:2010py,Liu:2017qah,ZhanduoTang:2023tsd}.

\section{T-matrix Approach}
\label{sec_TM}
The thermodynamic $T$-matrix provides a self-consistent quantum many-body formalism to compute 1- and 2-body correlation functions. As a potential-based approach (enabled by a suppressed energy-transfer in the scattering of massive particles), it utilizes a 3D reduced Bethe-Salpeter equation~\cite{Brockmann:1996xy} that is well suited for examining bound and scattering states in a strongly coupled environment~\cite{Liu:2017qah,ZhanduoTang:2023tsd}:                   
\begin{eqnarray}
\ T_{ij}^{L,a} ( z,p,p')&=&V_{ij}^{{L,a}} (p,p') +\frac{2}{\pi } \int_{0}^{\infty}k^{2}dk V_{ij}^{{L,a}} (p,k) G_{ij}^{0} (z,k)T_{ij}^{L,a} ( z,k,p') \ .
\end{eqnarray}
Here, $V_{ij}^{L,a}$ denotes the in-medium potential between particle $i$ and $j$ in color ($a$) and angular-momentum ($L$) channels; $G_{i j}^{0}$ is the 2-particle propagator, which is a convolution of 1-particle propagators; $p$ and $p'$ are the magnitudes of the incoming and outgoing momentum in the center-of-mass frame.
In the color-singlet channel, we use the ansatz $\widetilde{V}(r,T) = -\frac{4}{3} \alpha_{s} [\frac{e^{-m_{d} r}}{r} + m_{d}] -\frac{\sigma}{m_s} [e^{-m_{s} r-\left(c_{b} m_{s} r\right)^{2}}-1]$. The in-medium screening masses, $m_{d,s}$, 
and string breaking parameter, $c_b$, are determined from suitable constraints from thermal lQCD. 
In vacuum the potential reduces to the well-known Cornell form with the coupling constant, $\alpha_s=0.27$, and string tension, $\sigma=0.225$~GeV$^2$.

\section{Static Wilson Line Correlators from the $T$-matrix}
\label{sec_WLC} 
The Euclidean-time static WLC is related to the spectral function $\rho_{Q\bar{Q}}$ of a static $Q\bar{Q}$ pair via a Laplace transform,
\begin{equation}
W\left (r,\tau,T  \right )=\int_{-\infty}^{\infty}dE e^{-E \tau}\rho_{Q\bar{Q}}\left ( E,r,T \right ) \ ;
\label{wlc}
\end{equation}
$r$ is the distance between $Q$ and $\bar{Q}$ and $E$ their total energy relative to the HQ threshold, $2M_Q^0$ (numerically, $2\times 10^4$~GeV). The $Q\bar{Q}$ spectral function follows from the $T$-matrix as~\cite{Liu:2017qah,ZhanduoTang:2023ewm}
\begin{equation}
\rho_{Q\bar{Q}}\left ( E,r,T \right )=\frac{-1}{\pi}\mathrm{Im}\left [ \frac{1}{E-\widetilde{V}(r,T)-\Phi(r,T)\Sigma_{Q\bar{Q}}(E,T)} \right ] 
 \ ,
\label{rho_QQ}
\end{equation}
where $\widetilde{V}(r,T)$ is the static in-medium potential introduced in Sec.~\ref{sec_TM}.  The two-body selfenergy, $\Sigma_{Q\bar{Q}}$ contains the single HQ selfenergies and an $r$-dependent interference function~\cite{Liu:2017qah,ZhanduoTang:2023ewm}, $\Phi(r,T)$, which mimics 3-body diagrams from thermal-parton scattering off $Q$ and $\bar{Q}$. The first-order cumulant of the WLCs is defined as $m_1(r, \tau, T)=-\partial_\tau \ln W(r, \tau, T)$~\cite{Bala:2021fkm}, which can be interpreted as an effective mass that is commonly used in lattice QCD.

\begin{figure}[h]
\centering
\includegraphics[width=0.97\textwidth]{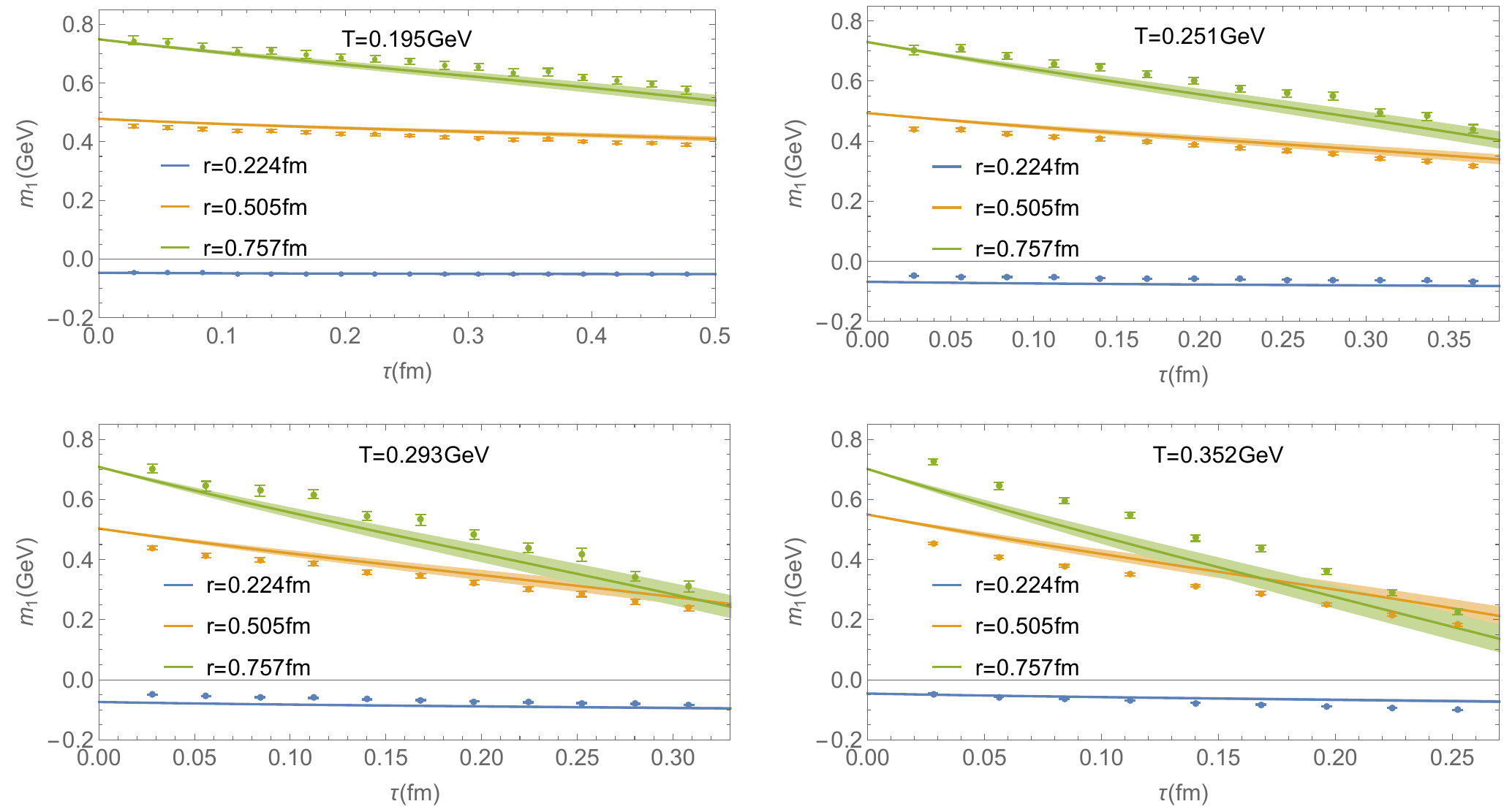}
\caption{$T$-matrix results for the first cumulant of static euclidean-time WLCs (lines) for three distances (color coded) at different temperature in each panel, compared to pertinent 2+1-flavor lQCD data~\cite{Bala:2021fkm}.} 
\label{fig_m1} 
\end{figure}

Our goal is now to constrain the in-medium potentials by the HQ WLCs while maintaining a description of QGP equation of state within a self-consistent calculation~\cite{ZhanduoTang:2023ewm}. For simplicity, the interference function $\phi(r,T)$ is adapted from previous work~\cite{Liu:2017qah}, but allowing for a $\pm10\%$ variation. Figure~\ref{fig_m1} shows our fit results for the cumulants of the WLCs, showing reasonable agreement with the lQCD results, albeit with some deviations at higher temperatures.
Different from previous constraints based on HQ free energies~\cite{Liu:2017qah,ZhanduoTang:2023tsd,ZhanduoTang:2023vvo}, we find significantly less screening for the input potential, especially at small and intermediate distances, with the string-screening mass being constant at $m_s$=0.2\,GeV, while $m_d$ and $c_b$ increase less than before~\cite{ZhanduoTang:2023ewm}.  

\section{Heavy-Quark Transport Coefficients}
\label{sec_transport}
\begin{figure}[htbp]
\begin{minipage}[b]{1.0\linewidth}
\centering
\includegraphics[width=1.0\textwidth]{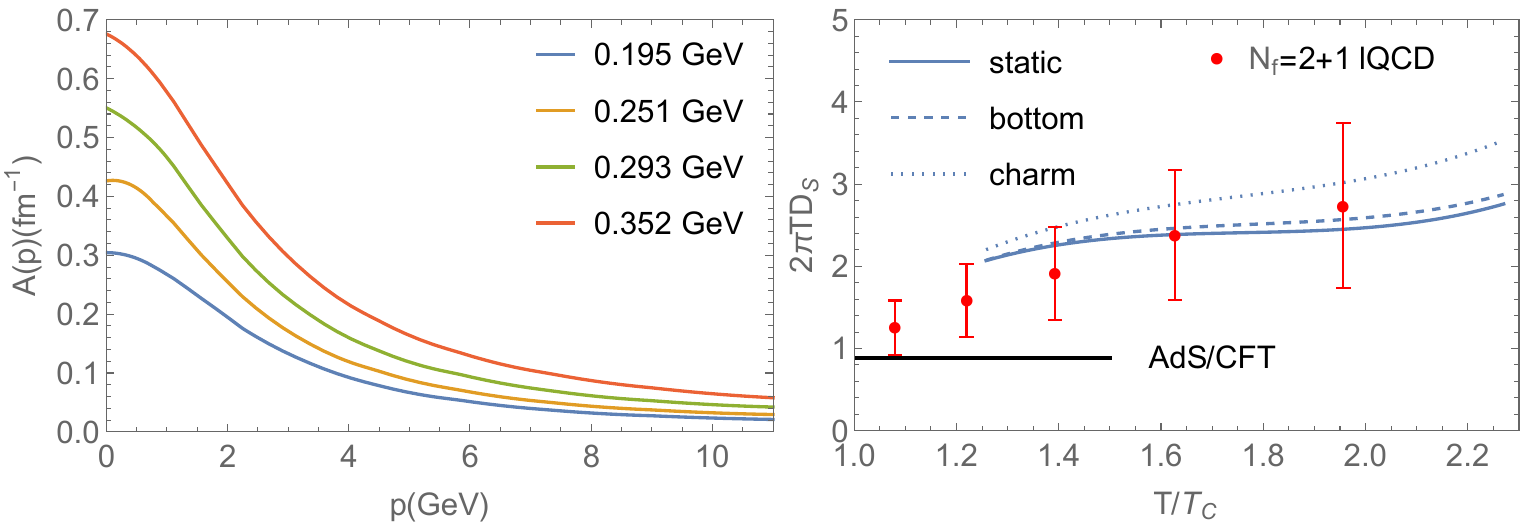}
\end{minipage}
\caption{Left: The charm-quark thermal relaxation rate as a function of charm-quark 3-momentum at different temperatures. Right: The spatial diffusion coefficient for static (blue solid line), bottom (blue dashed line) and charm (blue dotted line) quarks as a function of temperature (scaled by $T_c$=0.155 GeV) in comparison to the 2+1-flavor lQCD data~\cite{Altenkort:2023oms} (red dots with error bars) and the AdS/CFT estimate~\cite{Casalderrey-Solana:2006fio} (black line). The bare quark masses for charm and bottom are 1.359 and 4.681~GeV, respectively, and we approximate the static limit with a value of 10 GeV.} 
\label{fig_Ds}
\end{figure}

Next, we use the new potential to evaluate HQ transport properties in the QGP, specifically the thermal relaxation rate of charm quarks which can be expressed as 
$A(p)=\left \langle(1-\frac{\mathbf{p}\cdot\mathbf{p'}}{\mathbf{p}^2})\rho_{i}\rho_{i}\rho_{c}\right \rangle$; here, it is important to include off-shell integrations over the (broad) spectral functions, $\rho_{i,c}$, of the thermal-light and outgoing charm quarks to access the bound-state peaks below threshold~\cite{Liu:2018syc,Liu:2016ysz}.
The relaxation rate, $A(p;T)$, has a more gradual dependence on temperature than previously~\cite{Liu:2017qah,ZhanduoTang:2023tsd}, mostly due to larger values at the higher temperatures, see Fig.~\ref{fig_Ds} left. Consequently, the spatial diffusion coefficient, $D_s=T/(M_c A(p=0))$, scaled by $2\pi T$, has a flatter dependence on temperature, see Fig.~\ref{fig_Ds} right. In the static limit, the prediction are in reasonable agreement with recent lQCD data~\cite{Altenkort:2023oms}. 
The ramifications of these results on HF phenomenology observables in URHICs will be studied in the near future.

\section{Conclusions}
\label{sec_concl}
We have conducted a microscopic calculation of static $Q\bar Q$ Wilson line correlators employing the in-medium $T$-matrix formalism. With a moderately modified input potential, which features less screening at higher temperatures than previously, we have been able to obtain a fair description of pertinent lQCD data while maintaining a description of the EoS in the light-parton sector. This has been challenging for (resummed)  perturbative approaches and thus reiterates the prevalence of nonperturbative forces in the sQGP. When applied to HQ transport properties, we find that the increase in interaction strength at higher temperatures leads to a weaker temperature dependence for spatial HQ diffusion coefficient, which improved the agreement with recent 2+1-flavor lQCD results in the static limit. Applications of these results to the phenomenology of HF observables in heavy-ion collisions are in progress.

\section*{Acknowledgments}
This work has been supported by the U.S. National Science Foundation under grant no.~PHY-2209335 and by the U.S. Department of Energy, Office of Science, Office of Nuclear Physics through contract No. DE-SC0012704 and the Topical Collaboration in Nuclear Theory on \textit{Heavy-Flavor Theory (HEFTY) for QCD Matter} under award no.~DE-SC0023547.

%
%
%
%
%
\bibliography{refnew}
\end{document}